# Towards Real-time 3D Reconstruction using Consumer UAVs

Qiaosong Wang
University of Delaware, Newark, DE 19716, USA

## ABSTRACT

We present a near real-time solution for 3D reconstruction from aerial images captured by consumer UAVs. Our core idea is to simplify the multi-view stereo problem into a series of two-view stereo matching problems. Our method applies to UAVs equipped with only one camera and does not require special stereo capturing setups. We found that the neighboring two video frames taken by UAVs flying at a mid-to-high cruising altitude can be approximated as left and right views from a virtual stereo camera. Leveraging GPU-accelerated real-time stereo estimation and efficient P$n$P correspondence solving algorithms, our system simultaneously predicts scene geometry and camera position/orientation from the virtual stereo cameras. Also, this method allows user-selection of varying baseline lengths, which provides more flexibility given the trade-off between camera resolution, effective measuring distance, flight altitude and mapping accuracy. Our method outputs dense point clouds at a constant speed of 25 frames per second and is validated on a variety of real-world datasets with satisfactory results.

**Keywords:** 3D Reconstruction, Unmanned Aerial Vehicles, Stereo Matching, Pose estimation

# 1  INTRODUCTION

Recent advance in chipmaking and imaging sensor technologies enabled micro unmanned aerial vehicle (UAV) to reduce cost and quickly progress into consumer markets. Micro UAVs are widely used in agriculture, construction, transportation and film-making because of their exceptional stability, mobility and flexibility. Specifically, aerial mapping and photogrammetry provides a rapid, inexpensive, and highly automated approach to produce 3D digital assets for easy measurement, inspection, planning and management. However, current photogrammetry pipelines usually take hours to days to fully process captured imagery data, which greatly limits the ability for real-time data acquisition, onsite scan and inspection, emergency response and planning, etc. An ideal scanning pipeline should be able to map the scene in real-time as the drone travels along the flight path, and gives inspectors the ability to examine the partial scan even before the flight mission is finished. This could be potentially achieved by using high-end drones carrying aerial LiDAR systems. However, such systems are bulky, expensive, power-intensive and often unable to provide dense textured mesh. It is more desirable to enable real-time high-resolution mapping capabilities for consumer-level camera drones. Currently it is not possible to achieve this goal using photogrammetry approaches, for a few reasons. Firstly, photogrammetry approaches typically require global optimization for camera pose estimation, dense multi-view stereo matching, meshing and texture mapping. So, at framework level it already prevents online real-time mapping in an incremental manner. Secondly, the structure from motion (SfM) and multi-view stereo (MVS) components in the pipeline are computationally intensive and very time-consuming. To tackle this problem, we propose an alternative approach. At the core of our design is to use high-efficiency algorithms with lowest computational cost at each and every step of the pipeline. Therefore, we choose to use Harris corner instead of SIFT or SURF descriptors for feature extraction. We adopt real-time Belief Propagation (BP) stereo matching on downsampled images instead of running large-scale full-resolution MVS matching. We also use the very fast EP$n$P algorithm instead of performing Bundle Adjustment (BA) as in many SfM pipelines. The resulting system outputs accurate dense point cloud in near real-time at a constant speed of 25 frames per second.

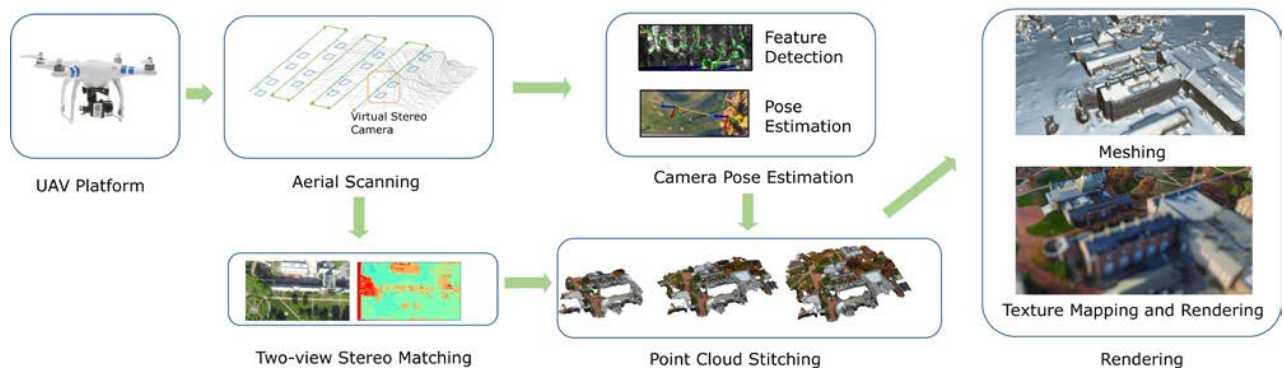

**Figure 1. The processing pipeline of our system.** We decompose the multi-view reconstruction problem into pair-wise two-view stereo matching problems. We perform fast camera pose estimation to stitch point clouds together and achieves near real-time speed at producing dense point clouds. Finally, we perform meshing, texture mapping routine to obtain the final scanned model and visualize in a web environment.

## 2 METHODOLOGY

### 2.1 CAMERA CALIBRATION

Our goal is to generate scans of buildings and terrain with accurate measurements. Therefore, it is very important to map camera measurements to real world coordinates. Since the GoPro camera we are using are wide-angle with severe distortion, we need to perform calibration to obtain intrinsic parameters of the camera. Our method supposes that near-by frames captured by UAVs flying at high altitudes can be approximated as left and right views from a generic stereo camera system. Therefore, for camera calibration we only recover the intrinsic parameters and ignores calibration of extrinsic parameters. We also assume that the camera has only radial distortion and does not have any tangential distortion. We use a planar checkerboard to calibrate the stereo camera. In OpenCV, this process can be done by calling *cvCalibrateCamera2()*. Once we obtained the calibration parameters, we could rectify and undistort individual images so that a point projected in the left image will always have its correspondent point projection laying on the same scan line in the right image. This rectification process is done by calling *cvInitUndistortMap()* and *cvRemap()*.

### 2.2 POINT CLOUD GENERATION

Once the images are rectified, we seek to obtain depth maps representing 3D coordinates of real-world points in the camera coordination system. Our GoPro camera device are with 2.7K resolution recording at 60 frames per second. It is almost impossible to run traditional global stereo matching methods on this resolution in real-time. To address this problem, we choose to perform stereo matching on low resolution image pairs first, and subsequently upsample the results with guidance from raw color images [3]. We use the GPU based Belief Propagation [4] and Joint Bilateral Upsampling [5] in our implementation. We downsample the stereo image pairs to the resolution of $320 \times 180$ and seek to minimize the following energy function [4]:

$$E_{\mathbf{X}}(d) = E_{D,\mathbf{X}}(d) + E_{S,\mathbf{X}}(d) = E_{D,\mathbf{X}}(d) + \sum_{\mathbf{Y} \in N(\mathbf{X})} M_{\mathbf{Y},\mathbf{X}}(d) \qquad (1)$$

Where $M_{\mathbf{Y},\mathbf{X}}(d)$ is the message vector passed from a pixel to its neighbor. $E_{D,\mathbf{X}}$ is the data term and $E_{S,\mathbf{X}}$ is the smoothness term. After a given number of optimization iterations, the label $d$ that minimizes $E_{\mathbf{X}}(d)$ is assigned to each pixel to form the disparity map. Next, we upsample the disparity map to obtain a high-resolution copy $D_p$ [5]:

$$D_p = \frac{1}{K_p} \sum_{q_\downarrow \in W} D'_{q_\downarrow} s(\| p_\downarrow - q_\downarrow \|) g(\| I_p - I_q \|) \qquad (2)$$

Where $p$ and $q$ be two pixels on the 2.7K full-resolution image $I$, $p_\downarrow$ and $q_\downarrow$ are corresponding pixels on the low resolution disparity map $D'$, $s$ is the spatial filter kernel, $W$ is the spatial support of kernel $s$, $g$ is the range filter kernel, and $K_p$ is the normalizing factor. By utilizing the range filter kernel, we could combine appearance information with fine levels of details from the full resolution image and spatial information from the downsampled disparity maps. Suppose the baseline of the camera is $B$, $du$ and $dv$ are lengths of the pixels on the horizontal and vertical axis, $f$ is the focal length of the camera, $(x, y, z)$ is the 3D point coordinate in the camera coordinate system and $(u, v)$ is the 2D projection of the point on the imaging plane, we could get:

$$[u, v, d] = \frac{1}{z}[\frac{fx}{du}, \frac{fy}{dv}, \frac{fB}{du}] + [u_0, v_0, 0]$$
$$[x, y, z] = \frac{B}{d}[u - u_0, \frac{(v - v_0)dv}{du}, \frac{fB}{du}] \quad (3)$$

Therefore, we could obtain point clouds for all virtual stereo image pairs with respect to their own camera coordinate systems.

## 2.3 POINT CLOUD STICHING AND MESHING

Our approach supposes near-by frames taken from the aerial video with a fixed interval can be treated as a set of left-right image pairs taken by virtual stereo cameras with identical baselines. However, it is worth noting that this assumption only holds locally. If the time difference between the two frames are too much, then we cannot assume ideal stereo camera setup (exact same camera pose) and must re-estimate camera pose and orientation. In this section, we discuss how to recover orientations and poses for all virtual stereo cameras. Firstly, we need to introduce a feature descriptor to build correspondence between two camera frames. Instead of commonly used SIFT/SURF feature descriptor, we choose to use the very fast Harris corner detector for our application. Suppose the image to be $I_{u,v}$ and we choose a patch with a small shift $[x, y]$, the change of intensity could be formulated as [1]:

$$E_k = \sum_{u,v} w_{u,v}\left[I_{x+u, y+v} - I_{u,v}\right]^2 = \sum_{u,v} w_{u,v}\left[Ax + By + C(x^2, y^2)\right]^2 = (x, y)M(x, y)^T \quad (4)$$

Where $A$ and $B$ are first order partial derivatives of the image, $w$ is the gaussian window function to improve robustness to noises. $M$ is a $2 \times 2$ matrix computed from the image derivatives. Harris corner is detected when the response $R$ reaches local maximum:

$$R = Det(M) - kTr^2(M) \quad (5)$$

The Harris corner is related to response from two directions and thus invariant to rotation and translation. Also, the computation is relatively simple and could be performed in real-time. For a given threshold $R$, we choose a patch centered at every Harris corner, and compute correspondence between frame $I_{t1}$ and $I_{t2}$, where $t$ is the timestamp. The compute of correspondence will only be performed within the given search range. We use Squared Difference (ZSSD) to calculate the cost. Once the correspondences are obtained, we seek to obtain the rotation and translation matrices $[\mathbf{R}, \mathbf{T}]$ which describes the relative camera pose between time $t_1$ and $t_2$. We adopt the very fast algorithm described in [2] for determining the position and orientation of the cameras given its calibrated intrinsics and a set of correspondences. At the core of this algorithm is to express 3D points as a weighted sum of 4 control points, reducing the computational complexity to $O(n)$. Suppose $\mathbf{A}$ is the intrinsic matrix obtained from the previous section, $\mathbf{u}_i$ are a set of Harris corners representing 3D points $\{\mathbf{p}_i^c\}_{i=1,\ldots,n}$, we now have:

$$\forall i, w_i \begin{bmatrix} \mathbf{u}_i \\ 1 \end{bmatrix} = \mathbf{A}\mathbf{p}_i^c = \mathbf{K}\sum_{j=1}^{4}\alpha_{ij} \cdot \mathbf{c}_j^c \quad (6)$$

Where $w_i$ are projective parameters, $\{\mathbf{c}_j^c\}_{j=1,2,3,4}$ are control points in the camera coordinate system.

By solving this linear system, we could obtain the translation and rotation matrices which maps control point coordinates from real-world coordinate system to camera coordinate system:

$$\mathbf{c}_j^c = [\mathbf{R},\mathbf{T}]\mathbf{c}_j^w \quad (7)$$

Thus, we could obtain camera poses for every virtually captured stereo pair obtained by the UAV. We calculate poses for both left and right images and average them to improve accuracy. This also enables us to calculate real-time speed, travelled distances and turn angles of the UAV in real-world metrics without using any GPS information. Combined with the point clouds we obtained from the previous section, we could stitch the point cloud together into a combined point cloud of the captured scene along the flight trajectory. We use the $[\mathbf{R},\mathbf{T}]$ matrices as initial poses and use the CUDA accelerated Iterative Closest Point (ICP) [6] algorithm to fine-register all point clouds together. We run ICP on the point clouds calculated from low-resolution disparity maps using Equation 3. The final module which includes orientation/pose estimation and fine-registration are running at a constant speed of more than 100 Hz on a desktop equipped with a standalone GPU.

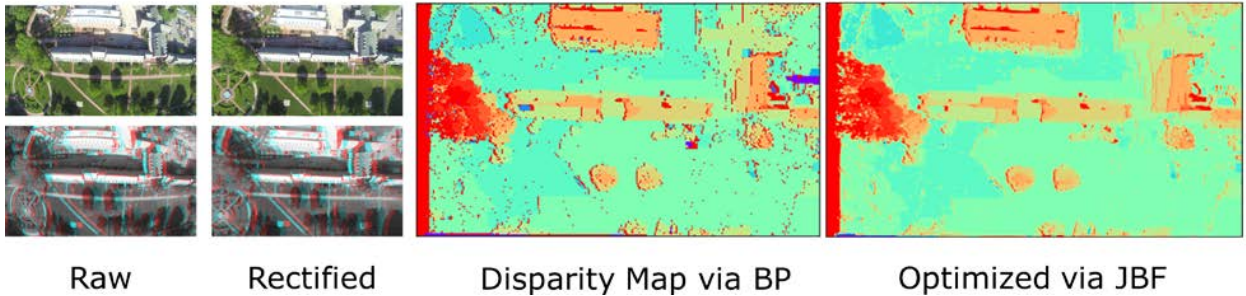

**Figure 2. Example output from out stereo matching module.** The first and second columns show raw images and rectified images and corresponding analygraphs. The third row shows the disparity map generated by belief propagation stereo matching. The final row shows optimized disparity map using joint bilateral upsampling. Note that small errors caused by BP is eliminated via JBF while all important edge details are preserved.

To this end, we have shown an efficient approach to generate dense point cloud data. We also leverage [7] for mesh reconstruction and [8] for texture mapping. Note that these two steps are used for generate visually-pleasing results and are not included in our real-time workflow. In the future, we will investigate different fast meshing and texture mapping approaches and work towards a more complete end-to-end 3D reconstruction pipeline.

# 3 EXPERIMENTS

## 3.1 HARDWARE

A DJI-Phantom Vision 2 Plus quadcopter with 2.4G datalink and iPad ground station is used in this project. It weighs about 1.2 kg with a diameter of 350 mm. The maximum forward, ascent and descent speed are 15 m/s, 6 m/s and 3 m/s, respectively. Automatic flights can be set up with a maximum altitude of 122 m (FAA 14 CFR part 107) and a maximum distance of 5 km. A GoPro Hero 3+ Black Edition camera is attached to a Zenmuse H3-3D gimbal to capture stabilized imagery off the UAV. The video signal is transmitted to the ground station via built-in Wi-Fi and an AVL58 FPV system is applied to increase the video transmission range to about 1 km in open space. An iOSD Mark II system is connected to the onboard controller to transmit real-time flight data such as power voltage, velocity, height, distance from the home point, horizontal attitude and GPS satellite number, etc. For all our experiments, we collect the videos, extract all frames and test on a desktop with NVIDIA Geforce Titan V 12GB graphics card. Although we process the frames in an offline fashion, there is a potential to run the same approach online with real-time speed. To achieve this, either a reliable high-speed datalink or an embedded onboard computing platform (e.g. NVIDIA Jetson) is required. Building such a system is outside of the scope of this work and thus we only evaluate on the offline video data.

## 3.2 SOFTWARE

We tested the CUDA accelerated BP algorithm with maximum iteration is set to 5 and maximum disparity value set to 16. We set both spatial and range filter size of the JBF to 15 and diameter of filtering neighborhood to 10 pixels. We divide the imaging plane to a $30 \times 30$ grid and calculate Harris corners inside each grid individually to prevent two many points detected at the same region given a hard threshold. We fly at 100 meters with the speed of 3 m/s and maintain altitude using the DJI built-in NAZA flight controller. We choose two frames out of every 10 frames to form a virtual stereo camera. The baseline is approximately 0.5 meters. We conducted experiments in various environments and upload to the results to [9]. We set field of view (FOV) to 60 degrees, physically based rendering (PBR) to "shadeless", and enable additional depth of field (DoF), bloom and Reinhard tone mapping effects. We also construct a dynamic web portal for navigation, preview and management of the uploaded 3D models. The finally results are shown in Figure 3.

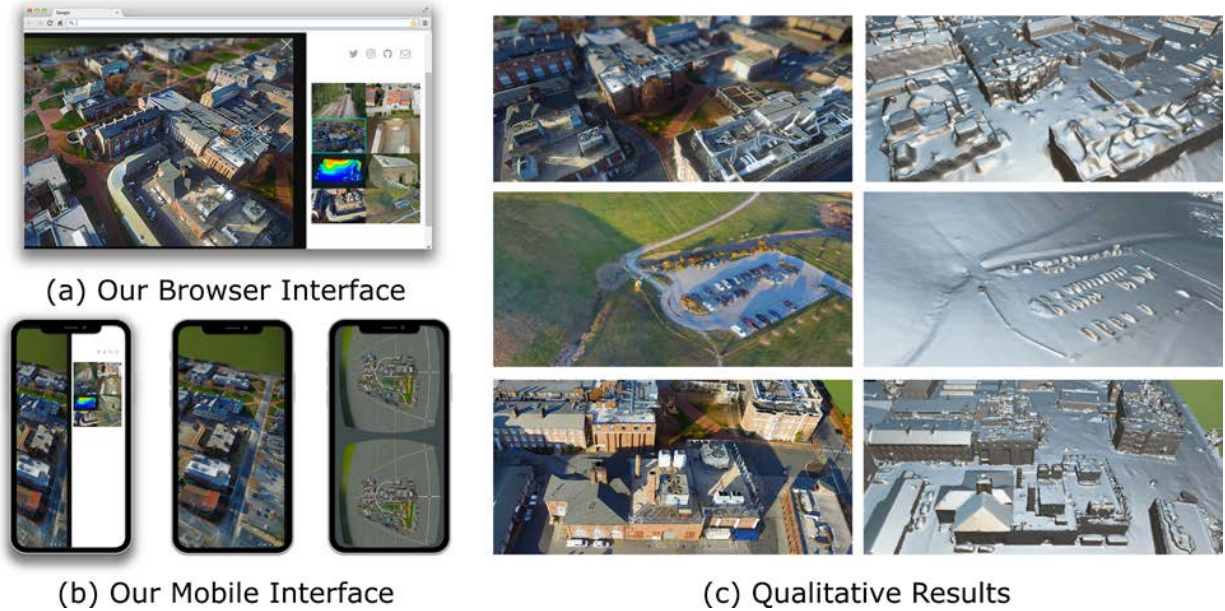

**Figure 3. Results.** Note that the VR view shown in (b) and the mesh render in (c) are provided by [9].

## 4 CONCLUSION

We have presented an efficient, near real-time solution for 3D reconstruction using off-the-shelf consumer UAVs. Our future efforts include comparing the proposed approach with other methods on 3D reconstruction benchmark datasets, and adding more global/loop-closure constraints to minimize trajectory errors.